\documentstyle[12pt]{article}
\begin{document}
\thispagestyle{empty}
\begin{titlepage}
\title{On the large order asymptotics of the wave function
perturbation theory}

\author{
   O.Yu. Shvedov  \\
{\small{\em Institute for Nuclear Research of the Russian
Academy of Sciences,  }}\\ {\small{\em 60-th October Anniversary Prospect
7a, Moscow 117312, Russia
}}\\ {\small and}\\ {\small{\em Moscow State University }}\\
{\small{\em Vorobievy gory, Moscow 119899, Russia}} }

\end{titlepage}
\maketitle

\begin{center}
{\bf Abstract}
\end{center}

The problem of finding the large order asymptotics for the
eigenfunction perturbation theory in quantum mechanics is studied.
The relation between the wave function argument $x$ and the number of
perturbation theory order $k$ that allows us to construct the asymptotics
by saddle-point technique is found: $x/k^{1/2}=const,
k\rightarrow\infty$. Classical eualidean solutions
starting from the classical vacuum play an important role in
constructing such asymptotics. The correspondence between the
trajectory end and the parameter $x/k^{1/2}$ is found. The obtained
results can be applied to the calculation of the main values of the
observables depending on $k$ in the $k$-th order of perturbation
theory at larges $k$ and, probably, to the multiparticle production
problem.

\newpage
It is well-known that the perturbation series in quantum mechanics, as well as
in quantum field theory, diverges \cite{D} because of the factorial growth
of the coefficients at large orders. High order asymptotics of the
coefficients,
which is important, for example, for the problem of the divergent series
summation,  has been discussed in many papers. For the quantities such as
the ground state energy in quantum mechanics \cite{BLGZJ,ZJ} or Green
functions in quantum field theory \cite{L}, the asymptotics is determined by
the euclidean finite action classical solution satisfying the following
boundary conditions: when ''euclidean time'' $\tau$ tends to $\pm\infty$,
the euclidean solution tends to the classical vacuum.

In this paper the role of classical euclidean solutions with other boundary
conditions is discussed. It occurs that when the large order asymptotics not
for eigenvalue, but for eigenfunction in quantum mechanics is considered,
one must treat the classical trajectory starting from the classical vacuum as
$\tau\rightarrow -\infty$ but finishing at some point $Q$ as $\tau=0$.

Namely, consider, as usual, the following dependence of the Hamiltonian on the
perturbation theory parameter $g$,
\begin{equation}\label{1}
H=-\frac{1}{2}\Delta + \frac{1}{g^{2}} V(gx),
\end{equation}
where $\Delta$ is the Laplace operator, $x=(x_1,...,x_n)$ is a set of
coordinates, function $V$ is a potential, which has a local minimum as its
argument is equal to zero. Without loss of generality, suppose that
$V(Q) \sim Q^2/2 + O(Q^3)$ as $Q\rightarrow 0$.The dependence (\ref{1}) was
considered in \cite{BLGZJ,ZJ}.

Consider the perturbation theory for the ground state eigenfunction. This
perturbation theory can be treated in several ways, namely, one can change the
variable $gx=Q$ and consider the wave function asymptotic expansion at
fixed $Q$; this expansion is a tunnel semiclassical expansion because of the
dependence (\ref{1}), the analog of the Planck constant is $g^2$. Large order
asymptotics of semiclassical expansion can be also found \cite{S1}, and the
asymptotics of the ground state energy perturbation theory, which was obtained
in \cite{BLGZJ,ZJ,BPZJ,RS} for various cases can be checked \cite{S1}.

Nevertheless, consider in this paper the ordinary way of treating the
eigenfunction perturbation theory,
\begin{equation}\label{2}
\Psi(x)=\sum_{k=0}^{\infty} g^k\Psi_k(x).
\end{equation}
The purpose is to find the $\Psi_k$ asymptotics at large $k$. In order to
do it, let us write $\Psi_k$ through the functional intrgral. Make use of
the path integral representation of the ground state wave function,
$$
\Psi(x)=\int Dq \exp \left( -\frac{1}{g^2}{\cal S}[gq]\right),
$$
where the integral is taken over the trajectories $q(\tau)$ satisfying
the boundary conditions, $q(-\infty)=0,q(0)=x,$ and ${\cal S}$
 is the euclidean
action of the theory,
\begin{equation}
{\cal S}[{\cal Q}]=\int d\tau [\dot{\cal Q}^2/2+V({\cal Q})].
\label{3}
\end{equation}
One can develop the perturbation theory for the ground state wave function
with the help of the expansion of the integrand in the path integral
representation into a series in $g$, the $k$-th order of this expansion can
be written as a contour integral, so the $k$-th order of the wave function
is given by
\begin{equation}
\Psi_k(x)=\int Dq \oint_{C(q)} \frac{dg}{2\pi i g^{k+1}}
\exp(-{\cal S}[gq]/g^2),
\label{4}
\end{equation}
the contour $C(q)$ runs around the origin counterclockwise.

Let us transform the integral (\ref{4}) to
 a saddle-point form. For this purpose rescale
the variables $q(\tau)$ and $g$ in $\sqrt{k}$ times:
\begin{equation}
g=\sqrt{\lambda}/\sqrt{k},q(\tau)=\xi(\tau)\sqrt{k}.
\label{5}
\end{equation}
This substitution means that the boundary condition for $q$,$q(0)=x$, must
be rescaled in $\sqrt{k}$ times, too. Thus, the $\Psi_k$ asymptotics at
large $k$ and fixed $x$ is not of a saddle-point type, while such asymptotics
as
\begin{equation}
k\rightarrow\infty, x/\sqrt{k}\rightarrow \xi_0=const
\label{6}
\end{equation}
 is
obtained from the rescaled integral with the help of saddle-point technique
and is given by the following expression,
\begin{equation}
\Psi_k(\xi_0\sqrt{k}) \sim \Gamma(k/2) \exp(-kA(\xi_0)),
\label{7}
\end{equation}
where the function $A$ equaling to
\begin{equation}
A(\xi_0)=\frac{1}{\lambda} {\cal S}[\sqrt{\lambda}\xi]
+\frac{1}{2}(\ln\frac{\lambda}{2}-1)
\label{7*}
\end{equation}
is taken in the extremum point, $\xi$ is a trajectory starting from
zero as $\tau\rightarrow-\infty$ and finishing at the point $\xi_0$ as
$\tau=0$. Varying the trajectory $\xi(\tau)$, one can obtain
that the function ${\cal Q}(\tau)=\xi(\tau)\sqrt{\lambda}$ satisfies the
classical euclidean equation of motion,
\begin{equation}
\ddot{\cal Q}=\nabla V({\cal Q}).
\label{8}
\end{equation}
 When one varies the quantity $\lambda$, the following
equation for it can be obtained:
\begin{equation}
\lambda/2=\int d\tau [V({\cal Q})-{\cal Q}\nabla V({\cal Q})/2],
\label{9}
\end{equation}
where $\tau \in (-\infty,0]$. We can treat the parameter (\ref{9}) as a
function of the end ${\cal Q}(0)=Q$ of the classical trajectory
${\cal Q}(\tau)$ satisfying eq.(\ref{8}). So we obtain the following relations
connecting the point $Q$ and the parameter $\xi_0$ in eq.(\ref{6}),
\begin{equation}
Q/\sqrt{\lambda(Q)}=\xi_0.
\label{10}
\end{equation}
When one searches for the asymptotic formula for $\Psi_k$ under conditions
(\ref{6}), one must first find the boundary condition for the classical
euclidean solution at $\tau=0$ from eq.(\ref{10}), then find the classical
trajectory arriving at this euclidean time moment in point $Q$ and, finally,
substitute the classical euclidean action on the trajectory and the
parameter $\lambda$ to eq.(\ref{7}). Notice that the function $\lambda$ in
eq.(\ref{10}) can be multivalued.

The correspondence (\ref{10}) is non-trivial and interesting. Namely, suppose
the trajectory ${\cal Q}(\tau)$ to spend all euclidean time in the
vicinity  of the point $Q=0$, so that the coordinates of the trajectory end
are small. Then the parameter $\lambda$ is of order $O(Q^3)$. Therefore, the
corresponding value of $\xi_0$ is large. Thus, if one makes an attempt to
find the behaviour of the asymptotics at large $\xi_0$, trajectories closed
to the path ${\cal Q}=0$ must be taken into account, so one is in need only
of knowledge of the first non-vanishing term of the difference $V-Q^2/2$
at small $Q$. This statement can be illustrate as follows. In the each order
of the perturbation theory the wave function is expressed as a product of the
$3k$-th order polynomial in $x$ function by the Gaussian exponent. It is
the term
of order $x^{3k}$ expressed only through the third derivatives
 $\partial^3 V/\partial Q_i \partial Q_j \partial Q_k$
of the potential $V$ at $Q=0$ that becomes the main one at large $\xi_0$.

We have seen that classical euclidean solutions starting at $\tau\rightarrow
-\infty$ from zero correspond to large order behaviour of $\Psi_k$, while
the parameter $\xi_0$ in the condition (\ref{6}) depends on the
coordinates of the trajectory end. We can consider all
such points in $Q$-space
that the discussed classical trajectory passes through them. According to
eq.(\ref{10}), each such point $Q$ can be associated with the corresponding
point in $\xi_0$-space by shifting of the parameter $\tau$ to a constant in
order to make the euclidean time moment at which the trajectory passes
through the point $Q$ to be equal to zero. Therefore, each trajectory in
$Q$-space is related with the trajectory in $\xi_0$-space (fig.1),
$\xi_0(\tau)={\cal Q}(\tau)/\sqrt{\lambda({\cal Q}(\tau))}.$ Focus the
attention on the distinction between this path and the trajectory
$\xi(\tau)={\cal Q}(\tau)/\sqrt{\lambda({\cal Q}(0))}.$ According
to the previous discussion, the trajectory  $\xi_0(\tau)$ starts from
infinity, while the trajectory $\xi(\tau)$ starts from the origin.
 Another interesting feature of the path $\xi_0(\tau)$ is that
the derivative of the function $A$ with respect to $\tau$ is written in the
form, $\dot{A}(\xi_0(\tau))=\pi_0\dot{\xi}_0$,where
$\pi_0=\dot{Q}/\sqrt{\lambda}$. The form is analogous to the formula $dS=PdQ$.

If the trajectory in $Q$-space finishes as $\tau\rightarrow +\infty$
at zero
 then the corresponding curve in $\xi_0$-space finishes in the point
$\xi_0=0$ as $\tau\rightarrow+\infty$, too, because the integral (\ref{9})
becomes the classical action at larges $\tau$, which is not equal to zero.
Thus, when one makes an attempt to find the asymptotics at small $\xi_0$,
one must consider the classical trajectory closed to the loop-like euclidean
solution (fig.1) with the action $S_0$. Since the function $\lambda$ is
approximately equal to $2S_0$ up to $O(Q^3)$, the function $A$ is equal to
$\frac{1}{2}\ln S_0 - \frac{Q^2}{4S_0}$ to the same accuracy. The $\Psi_k$
asymptotics as $k|\xi|^3 \ll 1, k|\xi|^2 \gg 1$ is, therefore, given by
$\Gamma(k/2)S_0^{-k/2}e^{k\xi^2/2}$. Notice that the pre-exponential factor
in the $\Psi_k$ asymptotics, which is omitted in this formula, diverges.
Namely, it is connected with the determinant of fluctuations near the
classical solution. When $Q$ is small and the trajectory is closed to the
loop-like solution, we come across the ''almost zero mode'' of the solution:
when we shift the parameter $\tau$ to a large constant, ${\cal Q}(0)$
is shifted to the small quantity. So the determinant is small.
Therefore, we can expect the following form of the $\Psi_k$ asymptotics at
fixed $x$ and large $k$,
\begin{equation}
\frac{\Gamma(k/2)}{S_0^{k/2}}\chi(x),
\label{10+}
\end{equation}
where $\chi(x)$ is not square integrable because of its exponential growth
at larges $x$, $\chi(x)\sim e^{x^2/2}$.

Since the ground state eigenvalue is defined by the behaviour of the
eigenfunction near the point $\xi_0=0$, large order asymptotics of
perturbation theory for it has the form similar to (\ref{10+}) up to a
pre-exponential factor. This formula coincides with the results obtained
in \cite{BLGZJ,BPZJ,ZJ,RS}.

Notice that the presented method can be applied to high order behaviour of
 perturbation theory not only for the ground state wave function but also
for the ground state density matrix equaling to the product
$\rho=\Psi(x)\Psi(y)$ because of the reality of the ground state wave function.
The density matrix and the $k$-th order for it can be presented in a form
analogous to (\ref{4}). We can change the variables like eq.(\ref{5}) and
find the $\rho_k$ asymptotics as
\begin{equation}
k\rightarrow\infty,x/\sqrt{k}\rightarrow\xi_1,y/\sqrt{k}\rightarrow \xi_2,
\label{11}
\end{equation}
which is given by
\begin{equation}
\rho_k(\xi_1\sqrt{k},\xi_2\sqrt{k})\sim
 \Gamma(k/2)\exp\left[-k\left(\frac{1}{\lambda}
(S[\sqrt{\lambda}\xi^{(1)}]+S[\sqrt{\lambda}\xi^{(2)}])+\frac{1}{2}\ln
\frac{\lambda}{2}-\frac{1}{2}\right)\right],
\label{12}
\end{equation}
where $\xi^{(1)}$ and $\xi^{(2)}$ are the trajectories satisfying the boundary
conditions $\xi^{(1)}(0)=\xi_1,\xi^{(2)}(0)=\xi_2$. We can also obtain that
the trajectories ${\cal Q}_1=\xi^{(1)}\sqrt{\lambda}$ and
${\cal Q}_2=\xi^{(2)}\sqrt{\lambda}$ are classical, while the parameter
$\lambda/2$ is equal to the sum of the integrals (\ref{9}) corresponding to
these two euclidean solutions.
 Notice that eq.(\ref{12}) can be also obtained by
making use of the $\Psi_k$ asymptotics under conditions (\ref{6}), but the
derivation is not so trivial, because one is in need of  estimation of
the sum $\sum_{n=0}^{k}\Psi_n(x)\Psi_{k-n}(y) $
under conditions (\ref{11}).

Consider the loop-like trajectory ${\cal Q}(\tau)$ shown in fig.1. Let
${\cal Q}_1(\tau)={\cal Q}(\tau-\tau_0),
{\cal Q}_2(\tau)={\cal Q}(\tau_0-\tau),\tau<0.$
This set of trajectories contributes to the asymptotics (\ref{12}) at
following values of $\xi_1,\xi_2$:$\xi_1=\xi_2={\cal Q}(0)/\sqrt{2S_0},$
while the parameter $\lambda$ is equal to $2S_0$, so that the asymptotics
(\ref{12}) at coinciding arguments
takes the form $\Gamma(k/2)S_0^{-k/2}$. Thus, we see that for
finding the density matrix large order asymptotics at coinciding arguments
which is necessary for obtaining the asymptotics of the main values of the
operators like $x_{i_1}...x_{i_m}$ (''Green functions'') one is in need of
making use of classical solution ${\cal Q}(\tau)$, and the corresponding
asymptotics has the form analogous to the ground state energy large order
behaviour. This result is in agreement with the technique developed in
\cite{L}.

But the results obtained in the presemnt paper allow us to find the asymptotic
behaviour of the $k$-th order of perturbation theory for the main values of
the observables $\hat{O}$ being the operators with the kernels $O(x,y,k)$
depending on $k$ as follows, $O(x,y,k)=f(x/\sqrt{k},y/\sqrt{k})
\exp(kg(x/\sqrt{k},y/\sqrt{k}))$. Namely, these quantities being the integrals
$\int dxdy\rho_k(x,y)O(x,y,k)$ can be approximately calculated by making use
of eq.(\ref{12}) and applying the Laplace method.

One can also calculate other quantities. Namely, the $k$-th order of
perturbation theory for the wave functions is presented as a product of a
polynomial function $\sum a_{i_1...i_n}x_1^{i_1}...x_n^{i_n}$ by the Gaussian
exponent $e^{-x^2/2}$. It occurs that the obtained result (\ref{7}) allows us
to find the $a_{i_1...i_n}$ asymptotics as $i_1/k,...,i_n/k$ tend to constants,
because these coefficients can be written through the contour integrals
having the saddle-point form under mentioned conditions.

The investigation of the wave function large order asymptotics gives us a
possibility to substitute it to perturbative recursive relations, to check
the exponential approximation (\ref{7}) and to find the pre-exponential
factor, as well as the corrections, not only by calculating the determinant
of fluctuations in the path integral approach, but also by direct analysis
of recursive relations, which allows us to consider the cases of singular
quantum corrections \cite{RS} to the Hamiltonian and of
instanton-anti-instanton pair \cite{BPZJ}. For semiclassical expansion
such analysis was made in \cite{S1}. From the recursive relations research
one can also confirm the asymptotics (\ref{10+}) and find the function
$\chi$ in it.

I hope that the suggested method of constructing asymptotics can clarify
the problem: what classical solutions contribute to the large orders of
perturbation theory and, in particular, find the correspondence between
the existence of this contribution and the parity of a number of negative
modes around the euclidean solution. The correspondence was found in
\cite{RS} for the ground state energy and confirmed in \cite{S1} for
the large orders of semiclassical expansion.

Probably, the suggested method can give rise to a new vision of a
 multiparticle production problem considered recently in many papers,
 see, for example, \cite{P1,P2,P3,P4}. For considering this problem it is
 important to know the behaviour of Feynman diagramms with a large
 number of the external lines and a large number of vertices, i.e. the
 main values of the observables depending on $k$ in the $k$-th order of
 perturbation theory at larges $k$ which has been already discussed.

The author is indebted to V.A.Rubakov for helpful discussions. This work
is supported in part by ISF, grant \# MKT000.

{\bf Figure caption}
Fig.1. Classical euclidean solution ${\cal Q}(\tau)$ and corresponding
curve in $\xi_0$-space.

\end{document}